\documentclass[fleqn,usenatbib,useAMS]{mnras}

\usepackage{graphicx}	
\usepackage{amsmath}	
\usepackage{amssymb}	
\usepackage{multicol}   
\usepackage{bm}		
\usepackage{pdflscape}	

\usepackage[T1]{fontenc}
\usepackage{ae,aecompl}
\newcommand{\Fig}[1]{Fig.~\ref{#1}}
\newcommand{\Figs}[2]{Figs.~\ref{#1} and \ref{#2}}
\newcommand{\Table}[1]{Table~\ref{#1}}

\title[Timing of the Sun's polar field reversals]{Probing the variations in the timing of the Sun's polar magnetic field reversals through observations and surface flux transport simulations}

\author[Golubeva et al.]{
Elena M. Golubeva,$^{1}$\thanks{E-mail: golubeva@mail.iszf.irk.ru}
Akash Biswas,$^{2}$\thanks{E-mail: akashbiswas.rs.phy20@iitbhu.ac.in}
Anna I. Khlystova,$^{1}$
Pawan Kumar,$^{2}$
Bidya Binay Karak$^{2}$
\\
$^{1}$Institute of Solar-Terrestrial Physics SB RAS, Lermontov Str. 126A, 664033, Irkutsk, Russia\\
$^{2}$Department of Physics, Indian Institute of Technology (Banaras Hindu University), Varanasi, 221005, UP, India
}

\date{Accepted XXX. Received YYY; in original form ZZZ}

\pubyear{2023}

\begin{document}
\label{firstpage}
\pagerange{\pageref{firstpage}--\pageref{lastpage}}
\maketitle

\begin{abstract}
The polar field reversal is a crucial process in the cyclic evolution of the large-scale magnetic field of the Sun. Various important characteristics of a solar cycle, such as its duration and strength, and also the cycle predictability, are determined by the polar field reversal time. While the regular measurements of solar magnetic field have been accumulated for more than half a century, there is no consensus in the heliophysics community concerning the interpretation of the Sun's polar field measurements and especially the determination of polar field reversal time. There exists a severe problem of non-reproducibility in the reported results even from studies of the same observational data set, and this causes an obstacle to make more accurate forecasts of solar cycle. Here, we analyze the solar magnetograms from four instruments for the last four cycles, to provide a more correct interpretation of the polar field observations and to find more accurate time of the reversals. We show the absence of triple (multipolar) reversals in Cycles 21\,--\,24, significant variations in the time interval between reversals in the hemispheres and in the time interval between a reversal and a cycle beginning. In order to understand the origin of the reversal time variation, we perform Surface Flux Transport (SFT) simulations and find out that the presence of the `anomalous' bipolar magnetic regions (BMRs) in different phases of a cycle can cause cycle-to-cycle variations of the reversal time within the similar range found in observations.
\end{abstract}


\section{Introduction}
\label{Sect-Intro} 
The strength of the large-scale magnetic field of the Sun oscillates while the polarity flips in about 11 years. Around the time of solar minimum, the magnetic field near the poles becomes maximum in its strength and then decreases with the progress of the solar cycle. Eventually, around the time of solar maximum, polar field reversal occurs. It is observed (and modelled by the surface flux transport (SFT) and dynamo models) that the decay and dispersals of the tilted Bipolar Magnetic Regions (BMRs) generates a poloidal magnetic field in low latitudes \citep{Babcock61,Leighton69}. This field is transported towards the poles through meridional circulation and cancels the opposite polarity fields that exist around the poles. Thus, the properties of the BMRs (primarily the tilt, amount of flux and the rate of emergence) and the surface flows determine the growth rate and the time of reversal of the polar field \citep{Bau04, JCS14, KMB18, Kumar22, Mordvinov22}. Recently, \citet{BKC22} explained how the strong nonlinear mechanism of toroidal flux loss due to BMR emergence causes the solar cycles decay in the similar way after the polar field reversal.

The polar magnetic field is a good measure of the strength of the next solar cycle \citep{Schatten78, Makarov89, Choudhuri07, Priyal14, Petrovay20, Kumar21}. This is because the poloidal magnetic field is transported down to the deeper convection zone where the differential rotation stretches it to induce a toroidal magnetic field which is the seed for the sunspots of the next cycle \citep{Karak14, Cha20}. If the polar field reverses early, then the next cycle begins early because the toroidal field and thus the sunspots for the next cycle will also emerge early. Furthermore, early polar field reversal will lead to early terminate the sunspot production and the ongoing cycle will reach to end early. This is because the early reversal will quickly annihilate the old polarity poloidal field, thereby stopping the generation of the toroidal field of the current cycle. Late polar reversal will cause the late end of the current cycle, resulting a strong asymmetry in its shape; the decline phase will be longer than the rising phase \citep{Kitchatinov18}. Moreover, a delayed polar field reversal will lead to the late beginning of the next solar cycle. Thus, in many ways, the timing of the reversals determines the length and shape of the current cycle and the beginning of the following cycle. Therefore, the analysis of regular long-term measurements of the magnetic field is one of the key points to understand the evolution of solar activity, to make its reliable predictions and to forecast the space weather and geomagnetic disturbances. 

In observations, we find that the polar field reversal timing is not fixed. It varies by a few years from cycle to cycle with respect to the average interval of 11 years. Thus, polar field reversals have attracted special attention since their discovery \citep{Babcock59} and promotion to the forefront in simulations of cyclic MHD-processes in the heliosphere \citep{Babcock61}. A reversal of the solar poloidal field completes global restructuring in the heliospheric magnetic configuration. Besides, it anticipates appearance of long-lived transequatorial coronal holes, which are the main geoeffective phenomenon during solar activity minima \citep{Petrie15}.

Experimental evaluations of the polar magnetic reversal time are distinguished by the particular ambiguity compared to the other temporal landmarks of the solar cycle. For example, in Cycle 24, some studies showed different polar reversal times in the range from March 2011 to October 2014 at the northern pole of the Sun, and in the range from March 2013 to April 2015 at the southern one \citep{Bertello15, Janardhan15, PastorYabar15, Sun15, Mordvinov16, Janardhan18, Mordvinov22}. Thus, the scatter of the measured reversal time is about 3.7/2.1 years in the north/south. In addition, individual authors noted the multiplicity of polarity reversals. Principally, a triple polar field reversal is not impossible for the Sun \citep{Mordvinov22}, but how often it may take place is the crucial question. 

Both the scatter of the measured polar field reversal time and the difference in descriptions of the physical process indicate the non-reproducibility of the reported results. A direct requirement of the scientific approach is to identify this problem and try to solve it.

Here, we consider this problem, concentrating our attention on the timing of the polar field reversals in cycles 21\,--\,24. From our analysis of the observational data, we find significant variations in the reversal time from one cycle to another. To probe the possible reason behind this variation, we perform SFT simulations and analyze how the variation of the BMR tilt properties and the presence of the `anomalous' active regions in the different phases of the solar cycles impact the timing of the reversals.

\section{Non-reproducibility in observations of the reversal time}
\label{Sect-PubResAnalysis}
\Table{T-simple} demonstrates some estimations of polar field reversal timing in Cycles 21\,--\,24. 
The presented results have been reported in ten chosen articles by the research groups who studied different observational data, using various methods.

\begin{table*} 
\setlength{\tabcolsep}{5pt} 
 \caption{Examples of polar magnetic reversal time estimations for Cycles 21\,--\,24. Here, T${_{\rm N}}$ and T${_{\rm S}}$ indicate the reversal dates reported for the Sun's North (N) and South (S) poles, correspondingly. The dates are shown as `YEAR/MONTH' or `YEAR/MONTH/DAY'. Angle b${_{0}}$ represents the heliographic latitude of the Sun's disk center as was observed from Earth on a date. The values of b${_{0}}$ correspond to the middle of a given day or the middle of a given month. Positive/negative values of b${_{0}}$ are favorable to observe the N/S pole of the Sun. The values of b${_{0}}$ may be considered as a guideline in assessing the accuracy of the reversal date estimations.}
 \label{T-simple}
\begin{tabular}{llllllrr}
\hline
Reference & Data Source & Approach & Latitude & T${_{\rm N}}$  & T${_{\rm S}}$  & b${_{0}}$(T${_{\rm N}}$) & b${_{0}}$(T${_{\rm S}}$) \\ 
\hline
\multicolumn{8}{c}{\textit{Cycle 21}} \\
\hline
\cite{Gopalswamy03} & NSO/KPVT & II & 70$^\circ$\,--\,90$^\circ$ & 1980/09\,--\, & 1979/07\,--\, & 7.2$^\circ$ & 4.3$^\circ$ \\
 & & & & \,--\,1982/07$^*$ & \,--\,1983/02$^*$ & 4.4$^\circ$ & $-6.8^\circ$  \\  
\cite{Ulrich13} & MWO/STT &  III & \quad 70$^\circ$ & 1980/05 & 1981/01 & $-1.3^\circ$ & $-3.1^\circ$  \\ 
\cite{Janardhan15} & NSO/KPVT & III & 45$^\circ$\,--\,78$^\circ$ & 1980/09 & 1979/03 & 7.2$^\circ$ & $-7.2^\circ$ \\
\cite{Janardhan18} & WSO, NSO/KPVT & I, III & 55$^\circ$\,--\,90$^\circ$ & 1981/06 & 1980/05 & 1.1$^\circ$ & $-2.5^\circ$ \\
\cite{Mordvinov22} & NSO/KPVT & III & 55$^\circ$\,--\,90$^\circ$ & 1982/06$^*$ & 1981/01 & 1.1$^\circ$ & $-4.6^\circ$ \\
\hline
\multicolumn{8}{c}{\textit{Cycle 22}} \\
\hline 
\cite{Wilson94} & MWO/STT & III & 50$^\circ$\,--\,85$^\circ$ & 1990/08\,--\, & 1989/06\,--\, & 6.8$^\circ$ & 0.2$^\circ$  \\
 & & & & \,--\,1991/07 & \,--\,1992/08 & 4.0$^\circ$ & 7.0$^\circ$ \\ 
\cite{Ulrich13} & MWO/STT & III & \quad 70$^\circ$ & 1990/03 & 1991/07 & $-7.2^\circ$ & 3.0$^\circ$ \\
\cite{Janardhan15} & NSO/KPVT & III & 45$^\circ$\,--\,78$^\circ$ & 1989/03 & 1990/03 & $-7.1^\circ$ & $-7.2^\circ$ \\
\cite{Janardhan18} & WSO, NSO/KPVT & I, III & 55$^\circ$\,--\,90$^\circ$ & 1991/04 & 1990/03 & $-5.6^\circ$ & $-7.2^\circ$ \\
\cite{Mordvinov22} & NSO/KPVT & III & 55$^\circ$\,--\,90$^\circ$ & 1991/08 & 1990/02 & 6.6$^\circ$ & $-6.8^\circ$ \\
\hline
\multicolumn{8}{c}{\textit{Cycle 23}} \\
\hline 
\cite{Gopalswamy03} & NSO/KPVT & II & 70$^\circ$\,--\,90$^\circ$ & 2000/03\,--\, & 2001/05\,--\, & $-7.1^\circ$ & $-2.6^\circ$ \\
 & & & & \,--\,2000/10$^*$ & \,--\,2001/10$^*$ & 5.8$^\circ$ & 5.8$^\circ$ \\
\cite{Ulrich13} & MWO/STT & III & \quad 70$^\circ$ & 2000/09 & 2000/09 & 7.2$^\circ$ & 7.2$^\circ$ \\ 
\cite{Janardhan15} & NSO/KPVT & III & 45$^\circ$\,--\,78$^\circ$ & 1999/09 & 2000/01 & 7.2$^\circ$ & $-4.5^\circ$ \\
\cite{Janardhan18} & WSO, NSO/KPVT & I, III & 55$^\circ$\,--\,90$^\circ$ & 2000/03 & 2000/06 & $-7.1^\circ$ & 1.1$^\circ$ \\
\cite{Mordvinov22} & NSO/KPVT & III & 55$^\circ$\,--\,90$^\circ$ & 2001/09 & 2001/10 & 7.2$^\circ$ & 5.8$^\circ$ \\
\hline
\multicolumn{8}{c}{\textit{Cycle 24}} \\
\hline 
\cite{Bertello15} & WSO & I & 55$^\circ$\,--\,90$^\circ$ & 2014/08/10$^*$ & 2013/07/26 & 6.4$^\circ$ & 5.3$^\circ$ \\
\quad \quad \quad \quad \quad \texttwelveudash & SOLIS/VSM & II & 60$^\circ$\,--\,75$^\circ$ & 2014/07/25$^*$ & 2013/08/15 & 5.2$^\circ$ & 6.7$^\circ$ \\
\cite{Sun15} & SoHO/MDI, SDO/HMI & II, III & 60$^\circ$\,--\,90$^\circ$ & 2012/11$^*$ & 2014/03 &  2.8$^\circ$ & $-7.2^\circ$ \\
\cite{Janardhan15} & SOLIS/VSM & III & 45$^\circ$\,--\,78$^\circ$ & 2011/03 & 2013/03 & $-7.2^\circ$ & $-7.2^\circ$ \\
\cite{PastorYabar15} & SDO/HMI & II & 70$^\circ$\,--\,80$^\circ$ & 2013/01/24 & 2014/02/28 &  $-5.4^\circ$ & $-7.2^\circ$ \\
\cite{Mordvinov16} & WSO & I & 55$^\circ$\,--\,90$^\circ$ & 2012/06/15\,--\,     & 2013/06/26 & 1.1$^\circ$ & 2.4$^\circ$ \\
 & & & & \,--\,2014/10/15$^*$ & & 5.8$^\circ$ & \\
\quad \quad \quad \quad \quad \texttwelveudash & SOLIS/VSM & II & 60$^\circ$\,--\,70$^\circ$ & 2012/05/14\,--\, & 2013/05/10 & $-2.7^\circ$ & $-3.2^\circ$ \\
 & & & & \,--\,2015/02/15$^*$ & & $-6.8^\circ$ & \\
\quad \quad \quad \quad \quad \texttwelveudash & SOLIS/VSM & II & 65$^\circ$\,--\,75$^\circ$ & 2012/12/31\,--\, & 2013/10/28 & $-3.0^\circ$ & 4.7$^\circ$ \\
 & & & & \,--\,2015/03/13$^*$ & & $-7.2^\circ$ & \\
\cite{Janardhan18} & WSO, SOLIS/VSM & I, III & 55$^\circ$\,--\,90$^\circ$ & 2012/06\,--\,& 2013/11 &  1.1$^\circ$ & 2.8$^\circ$ \\
 & & & & \,--\,2014/11$^*$ & & 2.8$^\circ$ & \\
\cite{Mordvinov22} & SOLIS/VSM & III & 55$^\circ$\,--\,90$^\circ$ & 2013/03 & 2015/04 & $-7.2^\circ$ & $-5.6^\circ$ \\
\hline  
\multicolumn{8}{c}{\textit{   }} \\
\multicolumn{8}{l}{MWO/STT -- \textit{Mount Wilson Observatory / Solar Tower Telescop}} \\ 
\multicolumn{8}{l}{NSO/KPVT -- \textit{National Solar Observatory / Kitt Peak Vacuum Telescope}} \\
\multicolumn{8}{l}{SDO/HMI -- 
\textit{Solar Dynamics Observatory / Helioseismic and Magnetic Imager}} \\ 
\multicolumn{8}{l}{SoHO/MDI -- 
\textit{Solar and Heliospheric Observatory / Michelson Doppler Imager}} \\ 
\multicolumn{8}{l}{SOLIS/VSM -- \textit{Synoptic Optical Long-term Investigations of the Sun / Vector SpectroMagnetograph}} \\
\multicolumn{8}{l}{WSO -- \textit{Wilcox Solar Observatory}} \\
\multicolumn{8}{c}{\textit{   }} \\
\multicolumn{8}{l}{  I -- Direct line-of-sight measurement of the polar field: time-series plot}\\
\multicolumn{8}{l}{ II -- Estimation of the polar field as a mean value in a selected area of magnetograms: a time-series plot, a butterfly diagram} \\
\multicolumn{8}{l}{III -- Estimation of the polar field as a mean value in a selected latitudinal zone of synoptic maps: a butterfly diagram} \\
\multicolumn{8}{c}{\textit{   }} \\
\multicolumn{8}{l}{$^*$ Authors reported unusual or multiple reversal of the polar magnetic field} \\  
\hline
\end{tabular}
\end{table*}

It is worth noting that some authors considered different latitudinal ranges above $\pm45^\circ$. These values were taken as conditional rough boundaries between the toroidal and poloidal components of solar magnetic field in the northern and southern hemispheres, when \cite{Babcock55} reported the lowest filling with large-scale magnetic field at heliographic latitudes in the vicinity of $\pm45^\circ$. However, for confidence, they measured the polar fields at latitudes above $\pm55^\circ$.

There is an upper latitudinal boundary for direct routine observations of the Sun's polar regions. Heliographic latitudes above $\pm82.75^\circ$ are irregularly observed from Earth and near-Earth space because of the tilt of the ecliptic to the solar equator by $7.25^\circ$. Both poles are synchronously visible only twice a year (at the beginnings of June and December), when the line connecting the Sun's and Earth's centers intersects with the solar equator so that zero heliographic latitude is observed in the solar disk center ($b{_{0}}=0^\circ$). The rest of the time, one of the poles is out of sight. Thus, all temporary series of polar field measurements have alternating semi-annual lacks of polar data and shorter lags in the circumpolar areas depending on latitude above $\pm82.75^\circ$. This is a serious problem to assess a time of a magnetic field reversal which takes place at the north Pole in December-May or at the south Pole in June-November. Therefore, the polar field reversal is usually considered as a reversal of integrated magnetic field in chosen lower latitudinal ranges in the polar caps. When a reversal is regarded up to $90^\circ$ \citep{Gopalswamy03, Bertello15, Sun15, Mordvinov16, Janardhan18, Mordvinov22}, the polar area of missing data is filled, using extrapolation.

In the set of papers considered here, we have identified three approaches used to study the polar magnetic field evolution.

First approach (noted in \Table{T-simple} as I) is based on routine direct observations of the polar fields with a relatively large aperture. Such observations with aperture of 3{\arcmin} are made in Wilcox Solar Observatory (WSO) since May 31, 1976. In this case, authors analyze a pair of time series for the north and south polar caps \citep{Bertello15, Mordvinov16, Janardhan18}.

In the second approach (noted in \Table{T-simple} as II), daily full-disk solar magnetograms are considered. On each magnetogram, a lot of pixels is selected in a given latitudinal polar zone and a given range of Stonyhurst longitudes around the central meridian to estimate mean magnetic field value. The result is displayed as a plot or time-latitude (butterfly) diagram \citep{Gopalswamy03, PastorYabar15, Sun15, Mordvinov16}.

\begin{table*} 
\setlength{\tabcolsep}{5pt}
 \caption{Analyzed time-series of full-disk magnetograms. Data from the ground-based instruments of MWO/STT and SOLIS/VSM are considered to study Cycles 21\,--\,23 and 24, correspondingly. Data from satellite instruments -- SoHO/MDI and SDO/HMI -- are considered to study Cycles 23 and 24, correspondingly. Cadence `Daily' means daytime with fairly good visibility of the Sun in the sky. Our SoHO/MDI and SDO/HMI working data-sets are formed to contain only one observation per day, taken at the very beginning of the day.}
 \label{data}
\begin{tabular}{lcccccccc} 
\hline
Data origin & \multicolumn{2}{c}{Considered} & Observation & Spectral & Aperture, & Image & LoS-field & Version \\
instrument & \multicolumn{2}{c}{period of time,} & cadence & line, & & size, & noise, & \\
  & Years & CRs & & \AA & arc-sec & pix & G & \\
\hline
\multicolumn{9}{c}{\textit{   }} \\
MWO/STT $^1$ & 1976.18\,--\,1996.72 & 1639\,--\,1913 & `Daily' & Fe I 5250 & 12.5$\times$12.5 & 512$\times$512 & 1 & Current \\ 
\quad \quad \texttwelveudash & 1996.64\,--\,2009.04 & 1913\,--\,2078 &  \texttwelveudash & \texttwelveudash & \texttwelveudash & 340$\times$340 & \texttwelveudash & \texttwelveudash \\  
SOLIS/VSM $^2$ & 2008.97\,--\,2017.85 & 2078\,--\,2196 & `Daily' & Fe I 6301.5 & 1,125$\times$1.125 & 2048$\times$2048 & 3 & Level 3 \\
SoHO/MDI $^3$ & 1996.64\,--\,2009.04 & 1913\,--\,2078 & 96 min & Ni I 6768 & 2$\times$2 & 1024$\times$1024 & 20 & Level 1.8.2 \\
SDO/HMI $^4$ & 2010.03\,--\,2020.02 & 2095\,--\,2225 & 720 sec & Fe I 6173 & 0.5$\times$0.5 & 4096$\times$4096 & 6.3 & Level 1.5 \\            
\hline
\multicolumn{9}{c}{\textit{   }} \\
\multicolumn{9}{l}{$^1$ 
\cite{Howard81, Ulrich02, Ulrich13} -- 
\url{ftp://howard.astro.ucla.edu/pub/obs/fits/}} \\
\multicolumn{9}{l}{$^2$
\cite{Keller03, Pietarila13} -- 
\url{https://solis.nso.edu/pubkeep/v7g/}} \\
\multicolumn{9}{l}{$^3$ 
\cite{Scherrer95, Liu04} -- 
\url{http://jsoc.stanford.edu/MDI/MDI_Magnetograms.html}} \\
\multicolumn{9}{l}{$^4$  
\cite{Scherrer12, Liu2012} -- 
\url{http://jsoc.stanford.edu/HMI/Magnetograms.html}} \\
\hline
\end{tabular}
\end{table*}

In the third approach (noted in \Table{T-simple} as III), a set of ready Carrington synoptic magnetic maps (`synoptic magnetograms') is analyzed to construct a butterfly diagram showing long-duration solar surface motions. First of all, this approach is important to study general magnetic field reorganization in a cycle of solar activity mainly at the latitudes of active regions \citep{Ulrich05, Kitiashvili20, Liu22, Wang22}. Some researchers consider the polar field reversals as a part of the global reorganization process \citep{Janardhan15, Sun15, Janardhan18, Mordvinov22}.

\Table{T-simple} shows significant scatter in the reversal time found with approaches I\,--\,III. However, the results presented here allow us to draw the next general conclusions. First of all, the north-south asynchrony of the reversals attracts attention. \Table{T-simple} mostly shows that the south polar cap is in the lead in Cycles 21 and 22, the northern one -- in Cycles 23 and 24. With that, the time interval between the north and south polar field reversal is about several months in Cycle 23, and about a year or more in other cases. It is noteworthy that four out of seven research groups \citep{Bertello15,Sun15,Mordvinov16,Janardhan18} noted the unusual or multiple reversal of north polar field (in \Table{T-simple} marked with *) in Cycle 24.

Nonetheless, in the cited estimations of the polar reversal time, the reproducibility of the result is not obvious and physical opportunity of multiple reversals and their frequency are under the question.

\section{Observational data and approach}
\label{Sect-DataApproach}
To study polar field reversals in Cycles 21\,--\,24, we analyze time series of solar full-disk line-of-sight (LoS) magnetograms from four instruments (\Table{data}). Empty and low-quality magnetograms were rejected.

Time of a polar field reversal is usually defined as a time of zero crossing in temporal variations of the polar magnetic field. That is why accuracy in estimations of magnetographical zero-level is critical. Control of zero-level offset (`magnetic bias') is a problem in registration of solar magnetic fields, especially using telescopes with medium or higher spatial resolution \citep{Demidov17}. As a rule, such instrumental artifact is compensated (with different accuracy) by mathematical processing of each magnetogram. This applies to the magnetograms considered here \citep{Pietarila13}. The best zero-field offset compensation is noted for SDO/HMI, magnetograms of which are often used as a kind of standard. It is not known how this problem was solved for the MWO/STT magnetograms updated in 2018--2019. We estimated timing of the polar field reversals both for all original sets of magnetograms and for all of them after our zero-level offset correction. 

To determine the magnetic zero-level offset, we apply a widely used method that was proposed by \cite{Ulrich02}. Next Section presents the correcting approach in detail. Here, we make a remark about its correctness. When a dominated value of weak magnetic fields is not zero, this method may be not quite suitable. But since during a polar field reversal a value of the average polar field is closest to 0~G, this gives hope for the relative correctness of corresponding zero-offset and, consequently, of the resulted time of polarity reversal. A zero offset, defined in this way, may be appropriate to take into account, first of all, for MWO/STT magnetograms.

Our processing sequence for each working magnetogram is as follows. Firstly, in each pixel, a longitudinal magnetic field value is divided by cosine of corresponding heliocentric distance to convert it into a value of the radial field $B_r$. Then, we allocate working regions for consideration of polar fields, using following angular limits on the solar disk: Stonyhurst longitudes in the range of $\pm45^\circ$ from the central meridian, latitudes above $\pm45^\circ$, and heliocentrical distances within the limits of $87^\circ$. The used latitudinal ranges are consistent with the main integrated limits from the studies cited in \Table{T-simple}. And finally, we calculate mean magnetic field values in latitudinal ranges with the regular latitudinal step equal to 5$^\circ$ in the case of MWO/STT, and 1$^\circ$ in all other cases. The choice of a step is based on the fact that in the extreme northern or southern position on the solar disk, the aperture of MWO/STT, SoHO/MDI, SOLIS/VSM or SDO/HMI overlaps heliolatitudinal angle at least about 9.3$^\circ$, 3.7$^\circ$, 2.8$^\circ$ or 1.8$^\circ$, correspondingly. Thus, a distribution of zonally averaged radial field is calculated from each magnetogram. Its angular resolution we believe to be suitable for statistical studies of the polar zones.

In turn, for each Carrington Rotation (CR), a mean zonal distribution is obtained by averaging a set of such latitude-field distributions corresponding to magnetograms observed during a considered CR.

To observe polar field reversals in each considered solar cycle, we arrange the mean zonal magnetic field distributions successively (CR-by-CR) into time-latitude diagrams for the north and south poles. Used temporal limits of each diagram correspond to solar cycle timing that is presented by the Solar Influences Data analysis Center (SIDC) on the web-page (\url{https://www.sidc.be/silso/cyclesminmax}). 

From the time-latitude diagrams we obtain corresponding variations of the field averaged over latitudinal ranges $\pm$(65\,--\,80)$^\circ$. Here, we do not consider diagram points with an increased noise . A threshold value is chosen as a maximum magnetic field at collective latitudes $\pm$(65\,--\,70)$^\circ$ of each diagram.

It is important to note that we calculate the time-latitude diagrams and time-field plots, avoiding extrapolation and leaving the lacks of magnetic field data on the polar caps without fitting. Like way was used by \cite{PastorYabar15} to study relations between the rotation axis and main magnetic axis of the Sun in 2010\,--\,2015. Such the approach lets to minimize artifacts in the results. 

The tilt by $7.25^\circ$ between the ecliptic and the Sun's equator causes correlated changes in the north and south polar magnetic fields. To minimize the impact of the seasonal component on the results, we use the approaches adopted for time series. Thus, considering an obtained field variation at latitudes of $\pm$(65\,--\,80)$^\circ$, we calculate three central-moving averaged curves with a window of 13~CRs (year): for the initial variation, for variation after median smoothing by 3 points and by 5 points. The median filtering removes isolated peak values (noise of 'salt and pepper'). If the transition time of the three curves through 0~G differs by no more than 1~CR, we take it into account. Otherwise, the reversal time is considered as undefined.

\section{Correction of magnetograms for zero-field offset in the polar caps}
\label{Sect-ZeroOffsets}
The main idea of the method proposed by \cite{Ulrich02} is that the frequency of magnetic field values relating to solar disk pixels should be normally distributed, so an observed field-frequency distribution should be a quasi-Gaussian curve with its peak at 0~G. Mathematical fitting of the Gaussian equation to an observed quasi-Gaussian curve makes it possible to find the Gaussian's center corresponding to a zero-level offset magnitude. To make a correction, this offset magnitude must be subtracted from the measured field values.

Magnitude of zero-level offset changes depending on a solar disk position and in time \citep{Liu04}. Therefore, when studying the polar fields, it is advisable to look for zero-level offset within the working areas on the disk in each magnetogram. To estimate magnetic zero-level offset, we use the next limits: Stonyhurst longitudes in the range of $\pm45^\circ$ from the central meridian, latitudes above $\pm55^\circ$, and heliocentrical distances within the limits of $87^\circ$.

\begin{figure} 
 \centerline{\includegraphics[width=\columnwidth,clip=]{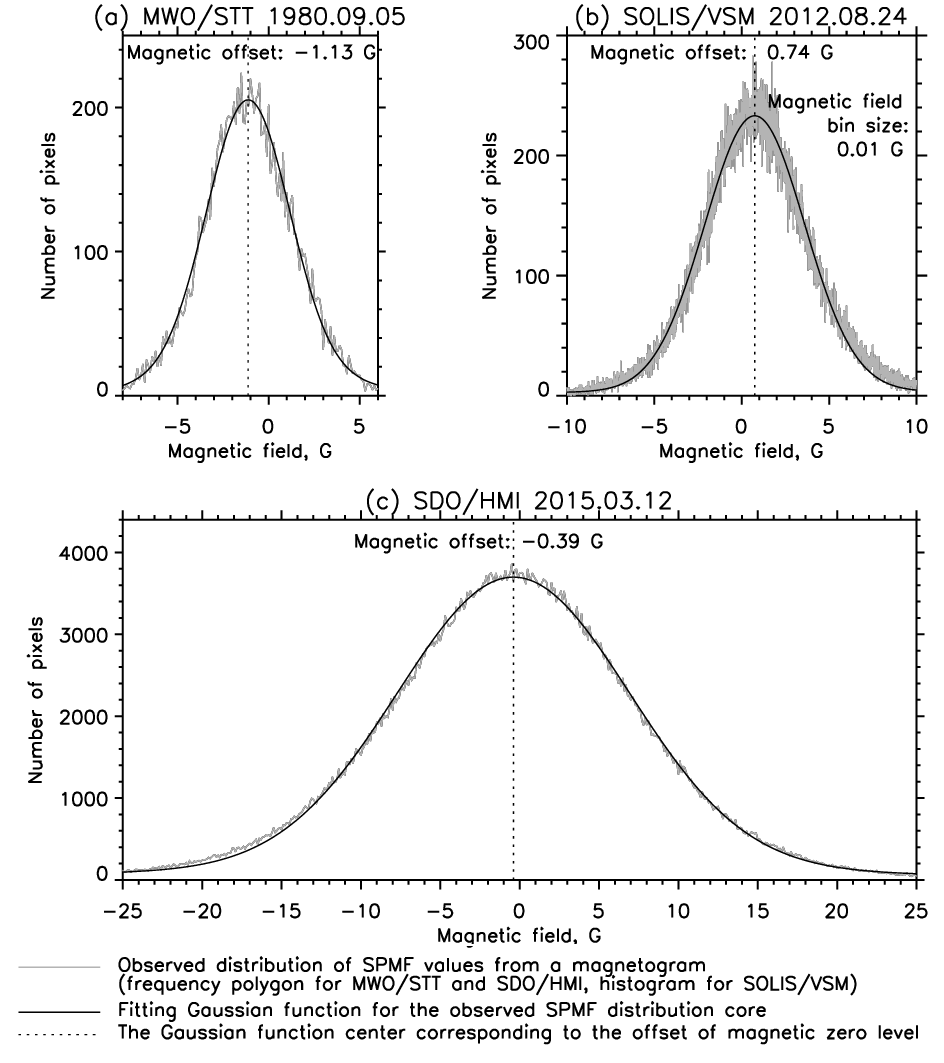} }
 \caption{Examples of Gaussian-fitting for the polar field measurements on magnetograms with classical field-frequency distributions. Magnetograms from MDI/STT (a) and SDO/HMI (c) are discrete numerical datasets, magnetograms from SOLIS/VSM (b) are continuous numerical datasets.} 
 \label{F-magoffset1}
\end{figure}

Magnetograms from MWO/STT, SOLIS/VSM, and SDO/HMI have quasi-Gaussian field-frequency distributions, so our approach to find their zero-level offsets is classical in these cases. Magnetic field values from a MWO/STT or SDO/HMI magnetoram are a discrete numerical dataset with the step of 0.01~G, so we find their observed quasi-Gaussian distribution as a frequency polygon. A magnetogram from SOLIS/VSM is a continuous numerical dataset, and we find its observed quasi-Gaussian distribution as a histogram with a bin size of 0.01~G. \Fig{F-magoffset1} shows examples of the observed field-frequency distributions, the corresponding mathematical Gaussian-fitting curves, and the found magnetic zero-level offsets for these three instruments.

\begin{figure} 
 \centerline{\includegraphics[width=\columnwidth,clip=]{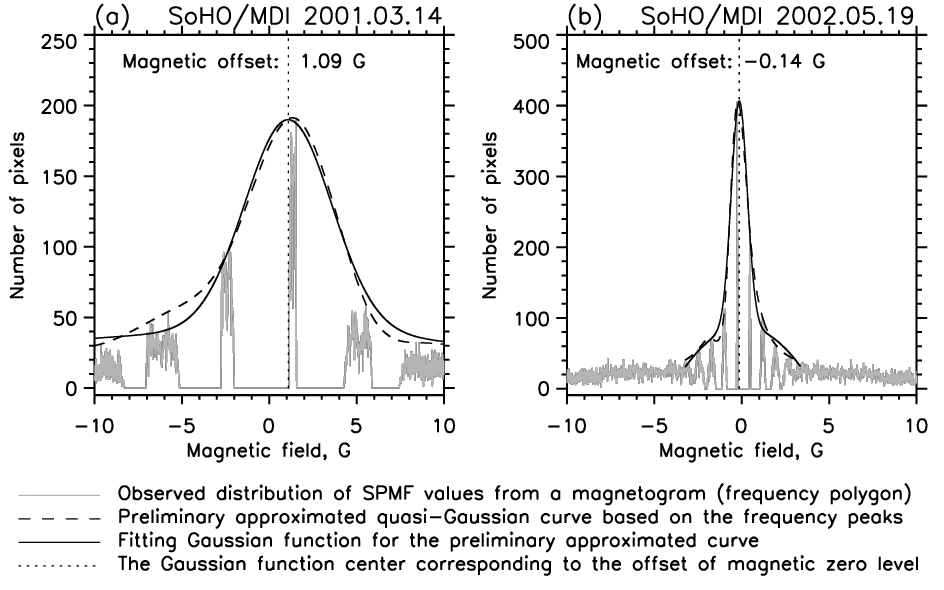} }
 \caption{Examples of Gaussian-fitting for polar field measurements from SoHO/MDI magnetograms. These numerical data-sets are discrete with the gaps near 0~G. In total, 98.96~\% of the considered magnetograms do not contain a measurement of 0~G. The size of a gap around 0~G is about 3.05 to 3.08~G (a) in 70.07~\% or about 0.61 to 0.62~G (b) in 29.93~\% of magnetograms.}
 \label{F-magoffset2}
\end{figure}

Field-frequency distributions of magnetograms from SoHO/MDI are not classical, their form looks like a `comb' with tips of `tooth' composing a Gaussian, and its numerical dataset is discrete with the step of 0.01~G and with a gap near 0~G. In this case, we perform spline-interpolation by spaced points of peaks in the `comb's teeth' to fill an observed field-frequency quasi-Gaussian curve and then do the final Gaussian-fitting to find zero-level offset. \Fig{F-magoffset2} shows examples of the observed field-frequency `combs', corresponding filled quasi-Gaussians, mathematically fitted Gaussian curves, and the resulting magnetic zero-level offsets for two kinds of magnetic field value distribution in magnetogram's numerical data sets.

\Fig{F-magoffset1} and \Fig{F-magoffset2} show examples of magnetograms with maximal absolute zero-level offsets observed during $\pm10$~CRs to a reversal time found in our study.

\begin{figure*} 
 \centerline{\includegraphics[width=1.\textwidth,clip=]{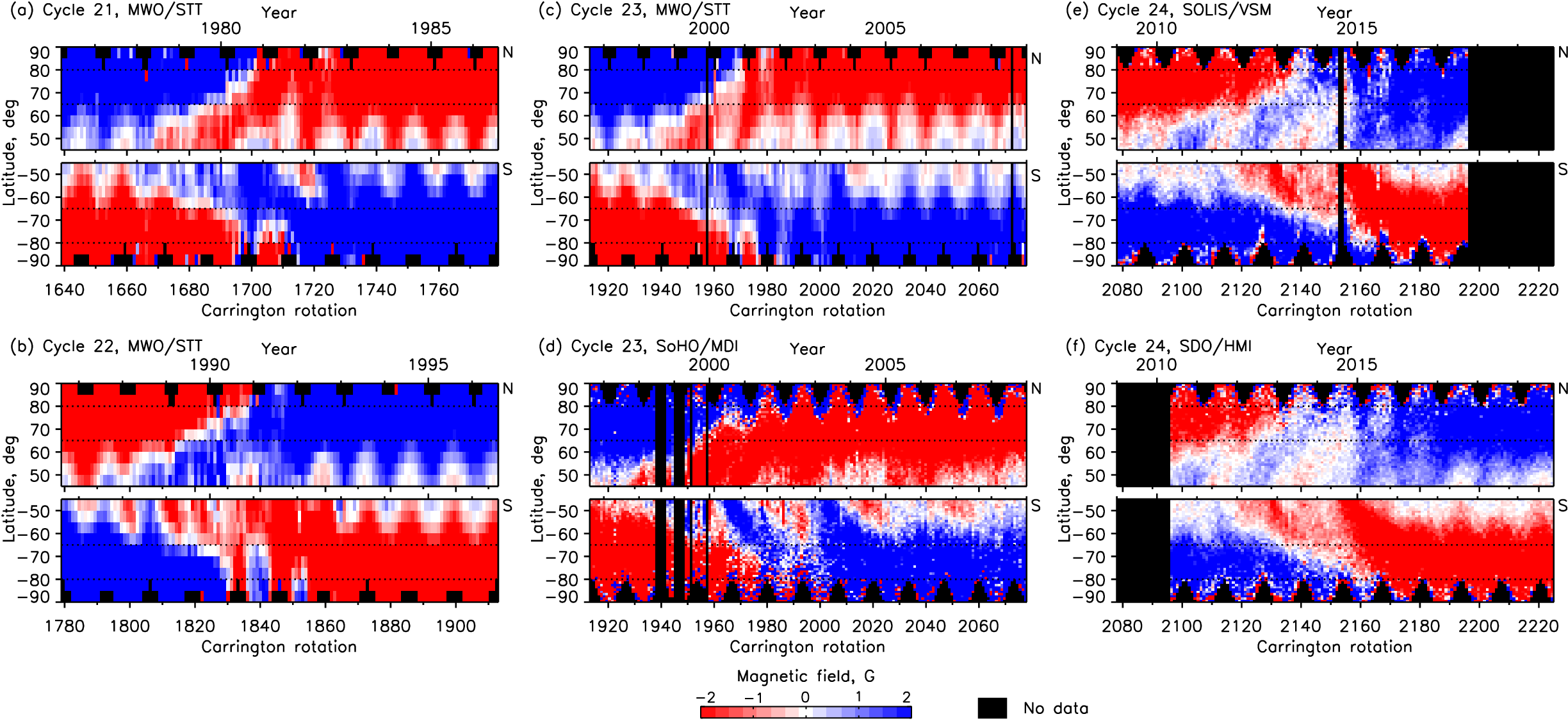} } 
 \caption{Time-latitude diagrams of solar magnetic fields on the north (N) and south (S) polar caps in Cycles 21\,--\,24. Here, magnetic fields are corrected for zero-level offset, but without correction for measurement errors. The field values greater/less than 2/-2~G are set to 2/-2~G to visualize reversals of the polar fields. Sampling period is 1~CR. Working central meridian distance is in the limits of $\pm45^\circ$. Latitudinal steps are 5$^\circ$ for MWO/STT (a\,--\,c) and 1$^\circ$ for others (d\,--\,f). The horizontal dotted lines show boundaries of circumpolar latitudinal range $\pm$(65\,--\,80)$^\circ$, where magnetic field variations may be unambiguously considered as poloidal component variations. At lower latitudes, a noticeable activity of the toroidal magnetic field is manifested. Higher latitudes have gaps in observations and high measurement noise near the limb.} 
 \label{diagrams}
\end{figure*}

The zero-level offset averaged per CR showed variations 
from $-0.73\pm0.02$~G to $0.71\pm0.02$~G for MWO/STT, 
from $-0.38\pm0.03$~G to $0.32\pm0.02$~G for SOLIS/VSM, from $-0.08\pm0.12$~G to $0.49\pm0.08$~G for SoHO/MDI, and 
from $-0.27\pm0.01$~G to $0.24\pm0.01$~G for SDO/HMI.
These statistics correspond to all cases with at least 9 magnetograms observed during a CR. It is important to note that in the case of the full disk on a magnetogram from SDO/HMI, the zero-level offset is 0~G, as a rule.

\section{Results from Analysis of Observations}
\label{Sect-Results}
\Fig{diagrams} shows the time-latitude diagrams of the polar fields in Cycles 21\,--\,24. Seasonal variations caused by the annual geometric effects are observed in the magnetic field measurements at all latitudes, but bringing the averaged field values to the threshold of $\pm$2~G makes them unobvious at polar latitudes. Below $\pm65^\circ$, periodic changes of magnetic polarity are evident. They are caused by low-latitude magnetic activity associated with the deep-seated toroidal fields. Above $\pm80^\circ$, magnetic field measurements are discontinuous in time. Therefore, it seems appropriate to consider the polar fields at circumpolar latitudes $\pm$(65\,--\,80)$^\circ$. \Fig{diagrams}(d\,--\,f) shows significant noise in the cases of SoHO/MDI, SOLIS/VSM and SDO/HMI, and it is in accordance with \Table{data}. Further, pixels with a high level of noise in the working latitude range are ignored.

\begin{figure*} 
 \centerline{\includegraphics[width=1.\textwidth,clip=]{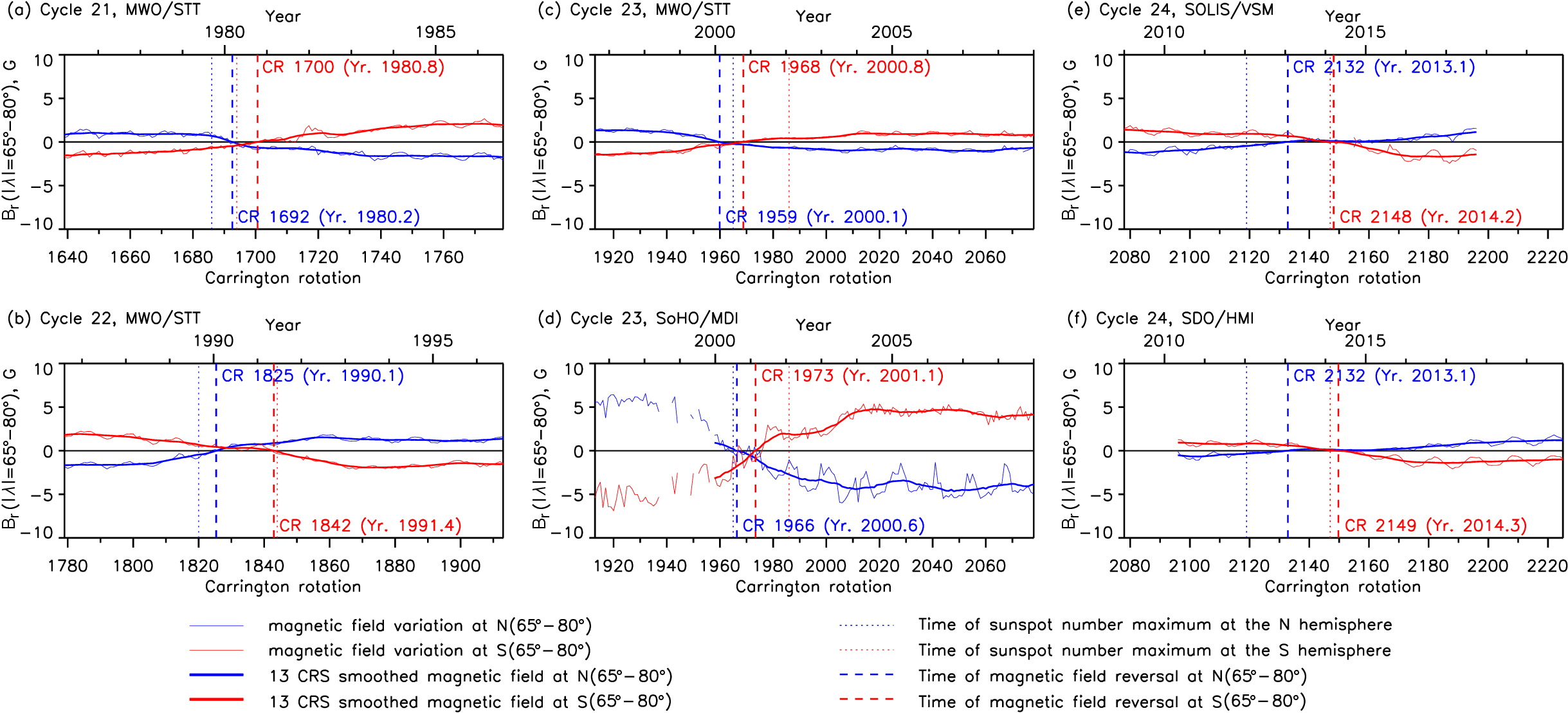} }
 \caption{Temporal variations of $B_r$ corrected for the noise and averaged over the regions limited by the central meridian distance of $\pm45^\circ$ and the latitudinal range $\pm$(65\,--\,80)$^\circ$ on the north (N) and south (S) polar caps with the time series step of 1~CR in Cycles 21\,--\,24.} 
 \label{plots1}
\end{figure*}

\begin{figure*} 
 \centerline{\includegraphics[width=1.\textwidth,clip=]{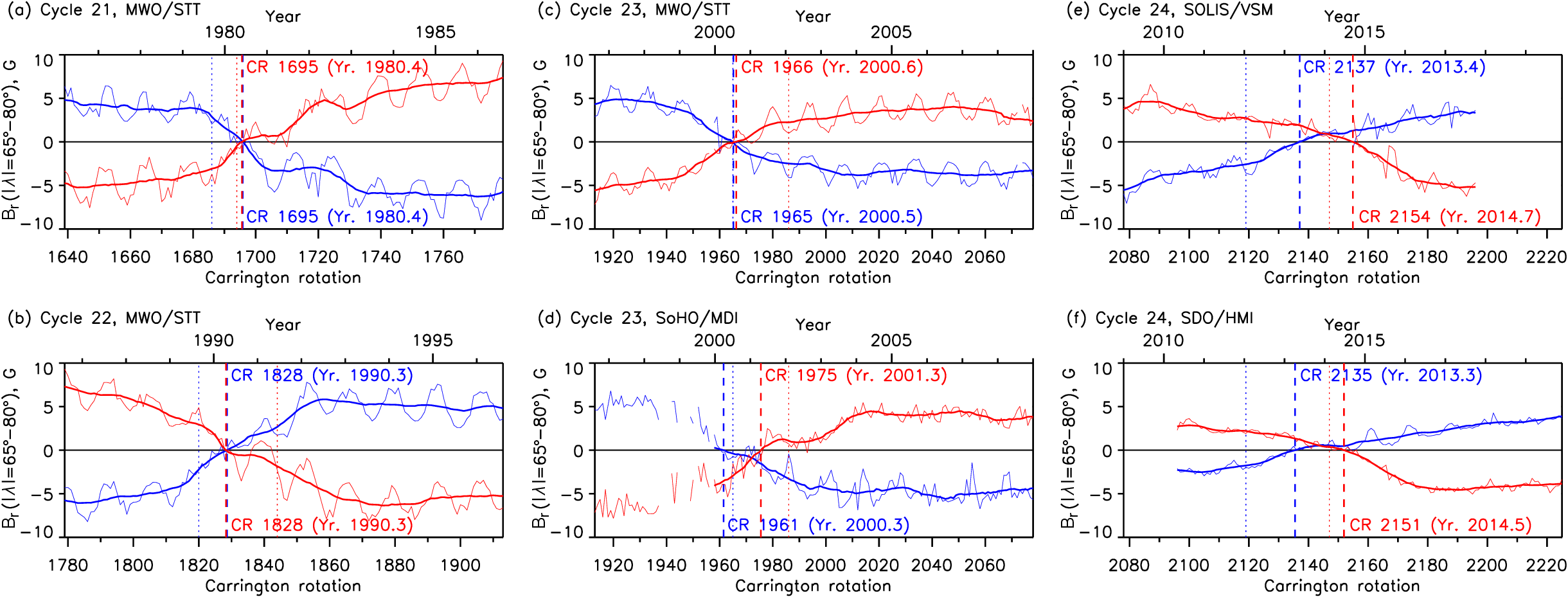} } 
 \caption{Temporal variations of $B_r$ corrected for zero-field offset and for the noise and averaged over the regions limited by the central meridian distance of $\pm45^\circ$ and the latitudinal range $\pm$(65\,--\,80)$^\circ$ on the north (N) and south (S) polar caps with the time series step of 1~CR in Cycles 21\,--\,24. Line designations are the same as in \Fig{plots1}.}
 \label{plots2}
\end{figure*}

For each considered cycle, we averaged the magnetic field measurements in the north or south latitude range. Then averaging over a moving window of 13~CRs was used to remove the seasonal geometric component. \Fig{plots1} shows the resulting variations and timing of the polar field reversal in the case of the original data, \Fig{plots2} -- in the case of data corrected for zero-level offset. Notably, in \Fig{plots1}(d), variations in the north and south unsmoothed polar fields of SoHO/MDI show the largest amplitudes compared to data from the other observatories and no correlation. Apparently, this is the result of significant mathematical processing of SoHO/MDI magnetograms during a series of recalibrations.

The following points are evident in the behavior of the curves in \Figs{plots1}{plots2}. First, the averaged radial magnetic field changes within $\pm10$~G. Second, in the periods of reversals, the unsmoothed field fluctuations take place near 0~G and show minimal amplitudes. This is favorable for the accuracy of zero-crossing time determination. 
Third, the smoothed polar fields retain their values near 0~G for a long time at the south in Cycles~21\,--\,23 and at the north in Cycle~24.
It is noteworthy that in the papers cited in \Table{T-simple}, the authors mainly noted an unusual or multiple reversion of the north polar field in Cycle 24 \citep{Bertello15, Sun15, Mordvinov16, Janardhan18}. Fourth, there were no multiple transitions through 0~G and the northern hemisphere was in the lead in the polar reversals in Cycles 21\,--\,24. This is consistent with the conclusions by \cite{Ulrich13} for Cycle~21, by \cite{Ulrich13,Janardhan15} for Cycles~22, by \cite{Ulrich13,Janardhan15,Janardhan18,Mordvinov22} for Cycles~23, by \cite{PastorYabar15,Janardhan15,Mordvinov22} for Cycle~24.

\Table{res} shows the timing of the polar field reversals as the results of averaging in three different ways (Section~\ref{Sect-DataApproach}). At latitudes $\pm$(65\,--\,80)$^\circ$, the time is defined for all the positions. The used ways show two neighbor CRs only for data from MWO/STT for Cycle~23 and SOLIS/VSM for Cycle~24. In all other cases, the convergence on the same CR takes place. It is important to note that the error of $\pm1$~CR, at least, is to be assumed here and for timing in \Table{T-simple}. 

\begin{table*} 
 \caption{Times of North/South (N/S) polar field reversals on latitudinal ranges of $\pm$(65\,--\,80)$^\circ$ and at three latitudinal levels in each of the ranges. These levels are the centers of the latitudinal ranges, where the magnetic field was initially averaged for the corresponding time-latitude diagrams (\Fig{diagrams}). The width of initial ranges is $5^\circ$ for MWO/STT and $1^\circ$ for the others. The sampling period is 1~CR. Limit of the central meridian distance is $\pm45^\circ$. Here, the estimations of reversal time are shown, the discrepancy of which does not exceed 1~CR. $T_r$ is temporal difference between a cycle beginning and a reversal of magnetic field on latitudes $\pm$(65\,--\,80)$^\circ$.}
\label{res}
\begin{tabular}{ccccccccc}
\hline
Data origin & Cycle & Start, & Maximum, & \multicolumn{4}{c}{Zero crossing time,} & $T_r$, \\ 
 & & CR$^*$  & CR(N/S)$^*$ & \multicolumn{4}{c}{CR} & CR(N/S) \\ 
\hline
 & & & & \multicolumn{4}{c}{\rm{Latitude}} & \\
 & & & & \rm{N/S} (65\,--\,80)$^\circ$ & \rm{N/S} 67.5$^\circ$ & \rm{N/S} 72.5$^\circ$ & \rm{N/S} 77.5$^\circ$ & \\           
\hline
 & & & & \multicolumn{5}{c}{\rm{Primordial data after the noise correction}} \\
\hline
MWO/STT & 21 & 1639 & 1686/1694 & 1692/1700 & 1690/1696 & 1693/1701 & 1694/1708 & 53/61 \\
\texttwelveudash & 22 & 1779 & 1820/1844 & 1825/1842 & 1820\,--\,21/1840 & 1825/1843 & 1828/1844 & 46/63 \\
\texttwelveudash & 23 & 1913 & 1965/1986 & 1959/1968 & 1956/1963 & 1960/1969 & 1964\,--\,65/1973 & 46/55 \\
SoHO/MDI & \texttwelveudash & \texttwelveudash & \texttwelveudash & 1966/1973 & - /1970 & 1972/1975 & - / - & 53/60 \\
SOLIS/VSM & 24 & 2078 & 2119/2147 & 2132/2148 & 2126\,--\,27/2141 & 2133/2148 & 2136/2154\,--\,55 & 54/70 \\
SDO/HMI & \texttwelveudash & \texttwelveudash & \texttwelveudash  & 2132/2149 &  - /2142 &  - /2151 & 2136/2155\,--\,56 & 54/71 \\
\hline
 & & & & \multicolumn{5}{c}{\rm{Data corrected for the noise and for zero-level offset}}\\
\hline
MWO/STT & 21 & 1639 & 1686/1694 & 1695/1695 & 1691/1692 & 1695/1696 & 1697/1704 & 56/56 \\
\texttwelveudash & 22 & 1779 & 1820/1844 & 1828/1828 & 1821/1824 & 1827/1829 & 1832/1830\,--\,31 & 49/49 \\
\texttwelveudash & 23 & 1913 & 1965/1986 & 1964\,--\,65/1966 & 1957/1956\,--\,57 & 1963/1966 & 1967/1970\,--\,71 & 51\,--\,52/53\\
SoHO/MDI & \texttwelveudash & \texttwelveudash & \texttwelveudash & 1962/1975 &  - /1972 & 1968/1976 &  - / - & 49/62 \\
SOLIS/VSM & 24 & 2078 & 2119/2147 & 2137/2152\,--\,53 & 2130/2142 & 2137/2150 & 2139/2164\,--\,65 & 61/74\,--\,75\\
SDO/HMI & \texttwelveudash & \texttwelveudash & \texttwelveudash & 2135/2151 &  - /2139\,--\,40 & 2138/2149\,--\,50 & 2135\,--\,36/2158\,--\,59 & 57/73 \\  
\hline
\multicolumn{8}{l}{$^*$ According to Monthly Hemispheric Sunspot Data from \url{https://www.sidc.be/silso/extheminum}} \\
\hline 
\end{tabular}
\end{table*}

Bearing in mind the presence of errors in the time points given in \Table{T-simple} and \Table{res}, we can note the previously published dates, which are closest to those found here. For north/south polar field reversals, \cite{Ulrich13} obtained CRs~1694\,--\,95/1703\,--\,04, 1826\,--\,27/1844\,--\,45, and 1967/1967 in Cycles~21\,--\,23, respectively, and \cite{PastorYabar15} -- CRs~2132/2147 in Cycle~24.

We can note the complete agreement between results by \cite{Ulrich13} and our ones for the MWO/STT in the reversals of the north fields with a correction for zero-level offset and the south fields without such a correction. This correction caused the time shift of +3~CRs for the north in Cycles~21\,--\,22 and the time shift of -2~CRs for the south in Cycle~23, and in these cases the time points by \cite{Ulrich13} are in the middle. Thus, results by \cite{Ulrich13} agree equally well with ours for the fields both with and without correction for zero-level offset. For Cycle 24, results by \cite{PastorYabar15} match with our ones obtained for the SDO/HMI's and SOLIS/VSM's fields without the offset correction. This is not surprising, since \cite{PastorYabar15} analyzed the same SDO/HMI data and used a method similar to ours. Note, our results obtained from SoHO/MDI data do not find strong confirmation in the results of the research groups considered here.

Of particular interest is the question about dynamics of the reversals. It is known that this process begins at the lowest latitudes of the polar caps and then drifts towards the poles. To estimate the drift in the operating latitudinal ranges $\pm$(65\,--\,80)$^\circ$, we chose three levels with $5^\circ$-spacing: $\pm67.5^\circ$, $\pm72.5^\circ$, and $\pm77.5^\circ$. The set of time points in \Table{res} lets us consider 18/19  time intervals between these latitudinal levels for the primordial/offset-corrected data, avoiding the time points for SoHO/MDI. As a result, in average, to get $5^\circ$ it takes about from 4.6$\pm$0.5 (primordial data) to 5.9$\pm$0.8~CRs (offset-corrected data). Therefore, in the assumption of the process uniformity, the drift from $\pm65^\circ$  to $\pm90^\circ$ requires about 1.5\,--\,2.5 years.

It is interesting to evaluate the difference in the dynamics of polarity reversals both between the latitude ranges $\pm$(67.5\,--\,72.5)$^\circ$ and $\pm$(72.5\,--\,77.5)$^\circ$, and between the polar caps. However, for this task, the considered samples are not statistically representative. Therefore, we can only note the qualitative trends. First, at the lower latitude range, the transit of a new magnetic polarity to a Pole of the Sun is slower. Second, this process is faster in the north polar cap.

\section{Results from the Surface Flux Transport simulations}
Here in this section, we would like to emphasize on the properties of the evolution of BMRs that determines the reversals in the polar fields and its eventual build-up. For this purpose we are utilizing the SFT model. The SFT model captures the essence of the so-called Babcock-Leighton (B-L) mechanism for the decay and dispersal of the tilted BMRs and the subsequent poleward motion of the remaining radial field under the influence of meridional circulation, differential rotation and horizontal diffusion \citep{WS89,Sheeley85,Bau04}. The central equation that the SFT models solve to simulate the above mentioned phenomena is the induction equation which in spherical geometry can be represented as follows:

\begin{eqnarray}
\frac{\partial {B_r}}{\partial t} = - \Omega(\lambda)\frac{\partial B_r}{\partial \phi} - \frac{1}{R_\odot \cos\lambda}\frac{\partial {}}{\partial \lambda} \left[ v(\lambda)B_r \cos\lambda \right]~~~~~~~~~~
\nonumber\\
+ \eta_H  \left[ \frac{1}{R_\odot ^2 \cos\lambda}\frac{\partial {}}{\partial \lambda}\left( \cos\lambda \frac{\partial {B_r}}{\partial \lambda} \right) + \frac{1}{R_\odot ^2 \cos^2\lambda}  \frac{\partial^2 {B_r}}{\partial \phi^2}  \right]
\nonumber\\
+ D(\eta_r) + S(\lambda, \phi, t)
 \label{eq:ind2}
\end{eqnarray} 

Here, $t$ , $\lambda$, and $\phi$ represent the time, latitudes and longitudes, respectively, $R_\odot$ is the solar radius, and $B_r$ is the surface radial field. 
The terms $v(\lambda)$ and $\Omega(\lambda)$ are the meridional circulation and the differential rotation on the solar surface which depend only on the solar latitude. $\eta_H$ and $\eta_r$ represent the horizontal and radial diffusivity respectively.
$S(\lambda,\phi,t)$ represents the source term of the radial field on the solar surface, and $D(\eta_r)$ captures the decay of the radial field due to radial diffusion. In this study, we have used the same SFT model which has been used in many previous studies such as \citet{Bau04,CJSS10}. Hence we do not discuss here regarding the profiles of the surface flows and the values of the different parameters in further detail. However, we note, this model uses the differential rotation profile provided by \citet{Snod83} (converted into synodic profile after taking into consideration the Earth's mean orbital angular velocity \citet{Skokic14})
which is of the following form:

\begin{equation}
 \Omega(\lambda) = 13.38 - 2.30 \sin^2\lambda - 1.62 \sin^4 \lambda \, \text{deg day}^{-1}
\end{equation}

\begin{figure} 
 \centerline{\includegraphics[width=0.55\textwidth,clip=]{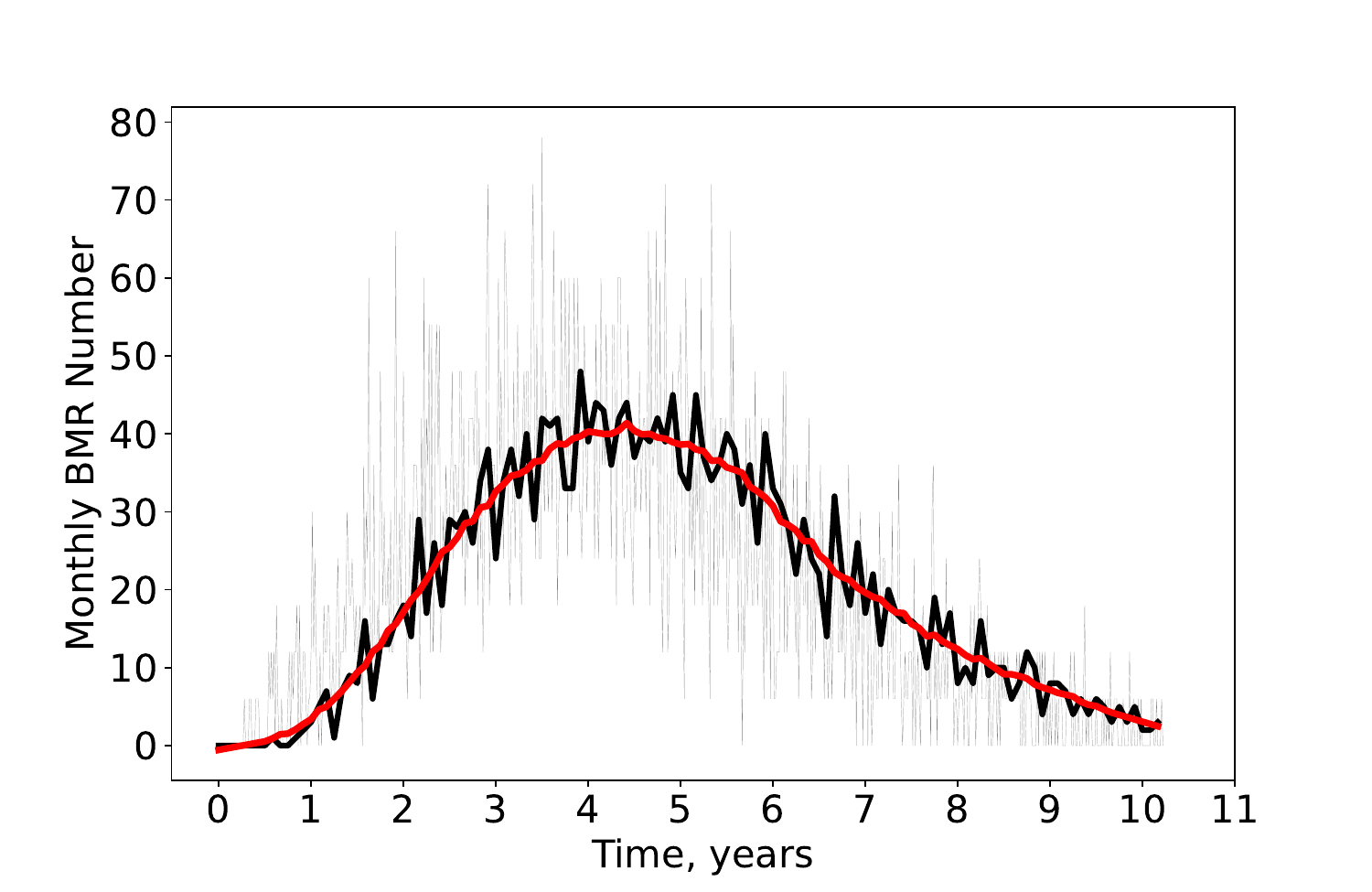} }
 \caption{A typical profile of one of the synthetic solar cycle in terms of monthly BMR number (the black curve and the corresponding smoothed red curve) overplotted on the five-day variations (the gray lines) of the BMRs.} \label{cycle_tilt}
\end{figure}

\begin{figure*} 
 \centerline{\includegraphics[width=1.\textwidth,clip=]{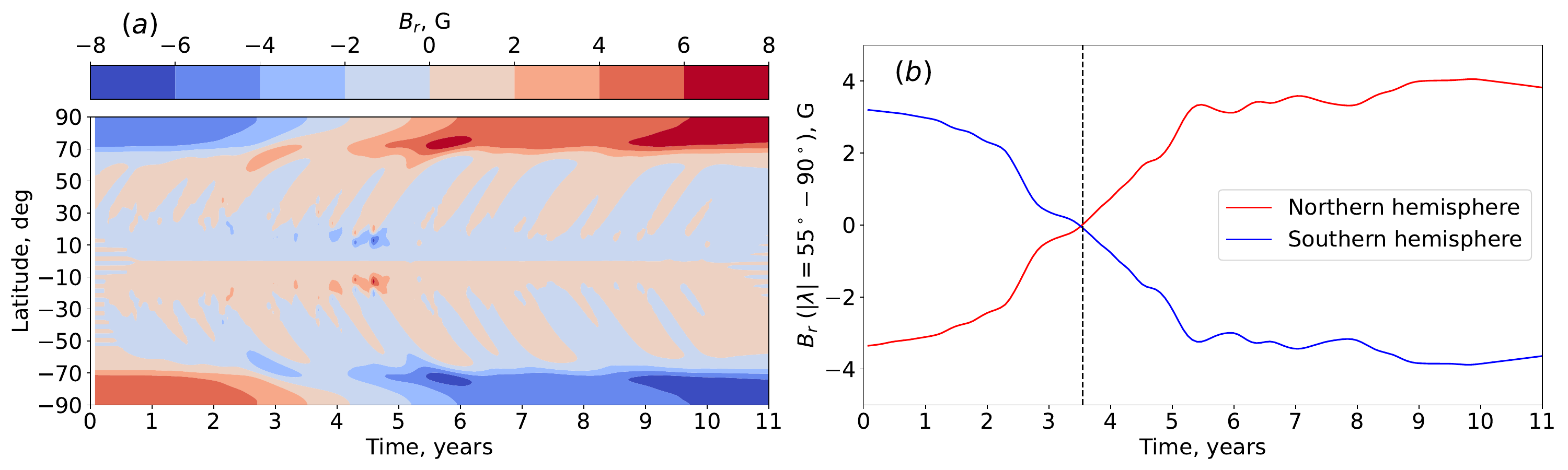} }
 \caption{Evolution of the radial surface field (a) and the corresponding evolution of the polar fields (b) from one of the simulations. The dashed vertical line in the panel (b) represents the reversal time $T_r$.} \label{polar_field}
\end{figure*} 

The various properties of the BMRs especially the distribution of the BMR tilt vary widely from one cycle to another and it is thought to be the primary reason for the wide range of variability in the long term modulation of the solar cycles \citep{KM17, LC17, BKUW23,Kar23}. The tilt of the BMRs increases with the increase in the emergence latitude according to the formula known as the Joy's law \citep{Hale19}. However it has been observed that, although the tilts obey the Joy's law statistically, they exhibit a huge scatter (\citet{How91, Fisher95, Jha20}, also see Fig.~4 of \citet{Kar23}). Due to this scatter in the BMR tilts, the contribution of different BMRs in the ultimate build-up of the polar field vary widely \citep{JCS14, KM18}. Often, the tilts and the orientation of magnetic polarities of certain BMRs are opposite to that of the regular BMRs, as a result, they contribute negatively in the polar field build-up. In certain cases, they can produce large fluctuations in the polar field leading to extreme events like the Maunder Minimum \citep{NLLPC17}. These kinds of BMRs are known as the `anomalous' or `rogue' BMRs (such as anti-Joy or anti-Hale BMRs). It is obvious that the presence of these anomalous BMRs will have an impact on the amplitude and reversal time of the polar field. However, there is a lack of understanding regarding how the polar field reversal time is affected due to the presence of these BMRs in different phases of the solar cycles, and there has been very few studies in the past attempting to qauntify their impact on reversal timing \citep{NLLPC17, Pal23}. 
We note that \cite{Pal23} included anomalous regions in varied amounts and with spatio-temporal variations 
 (having flux content of $5\%$ and $10\%$ of the total flux), claiming that the anti-Hale and anti-Joy regions impact the evolution of the polar field in a similar manner, however they did not include the significant observed scatter in BMR tilts around Joy's law \citep[for example, see Fig.~4 of][] {Kar23} in their study.
Here we utilize the aforementioned SFT simulations to take a look at how the variation in the BMR tilt properties and the inclusion of the anomalous BMRs (anti-Joy and anti-Hale) 
in varied amounts in
different phases of the solar cycles impact the timing of the polar field reversal. 
We also investigate the relative impact of anti-Joy and anti-Hale BMRs on reversal timing.
Here we would like to mention that, in this study we refer the BMRs as the  active-regions that tightly correlate with the solar cycle and strictly follow the time-latitude trend of the butterfly diagram. The ephemeral regions that do not tighly follow the solar cycle and have large scatter in their tilts  \citep{HST03, Anu23} are not included in our simulations.

To produce the synthetic spatiotemporal profiles of the BMRs that closely resembles the observed properties, we follow the analytical prescriptions as provided in \citet{HWR94} and \citet{Jiang18}. 
In the \Fig{cycle_tilt} we present the profile of the synthetic solar cycle in terms of monthly BMR number.
For the tilts of the BMRs, we introduce a Gaussian scatter around the values of tilts obtained from the Joy's law: $\gamma = \gamma_0\sin \lambda$,with $\gamma_0 = 35^\circ$ \citep{Hale19,WS89,How91}. 
The  scatter in BMR tilts around the Joy's law randomly varies from one cycle to another.

The `regular' or the Hale--Joy type BMRs have tilts within the range of $0^\circ< \gamma < 90^\circ$. 
The tilts of the anti-Joy BMRs are within the range of $-90^\circ< \gamma < 0^\circ$ whereas the tilts of the anti-Hale BMRs for which the conventional longitudinal orientation of the BMR polarities are flipped, falls within the range of $-180^\circ< \gamma < -90^\circ$. 
For our study, we allocate the amount of these anomalous BMRs randomly for different cycles. The percentage of the anti-Joy BMRs have been taken to be within the range of $10-30\%$ whereas the percentage of the anti-Hale BMRs have been taken within the range of $3-7\%$ keeping consistent with the observations \citep{McClintock+Norton+Li14, Jaramillo21}. 
These synthetic BMRs are the inputs to the SFT code for studying the evolution of the polar field.

To compute the strength of the polar field from the simulations, we produce magnetogram maps at each 27 days interval (period of the Bartels Rotation, i.e. the rotation period of Sun near the equator in our model). 
From these maps the longitudinal averages of the surface magnetic fields are taken which provides the latitudinal migration of the fields with time. In the next step, the average value of the radial surface magnetic field from $55^\circ$ to $90^\circ$ latitudes is taken to be the polar field strength in each of these maps.
The evolution of the surface radial field is shown in the panel (a) of \Fig{polar_field} whereas, the evolution of the corresponding polar field is presented in the panel (b), the vertical black dashed line shows the time of the reversal ($T_r$) of the polar field. 

\begin{figure*} 
 \centerline{\includegraphics[width=1.0\textwidth,clip=]{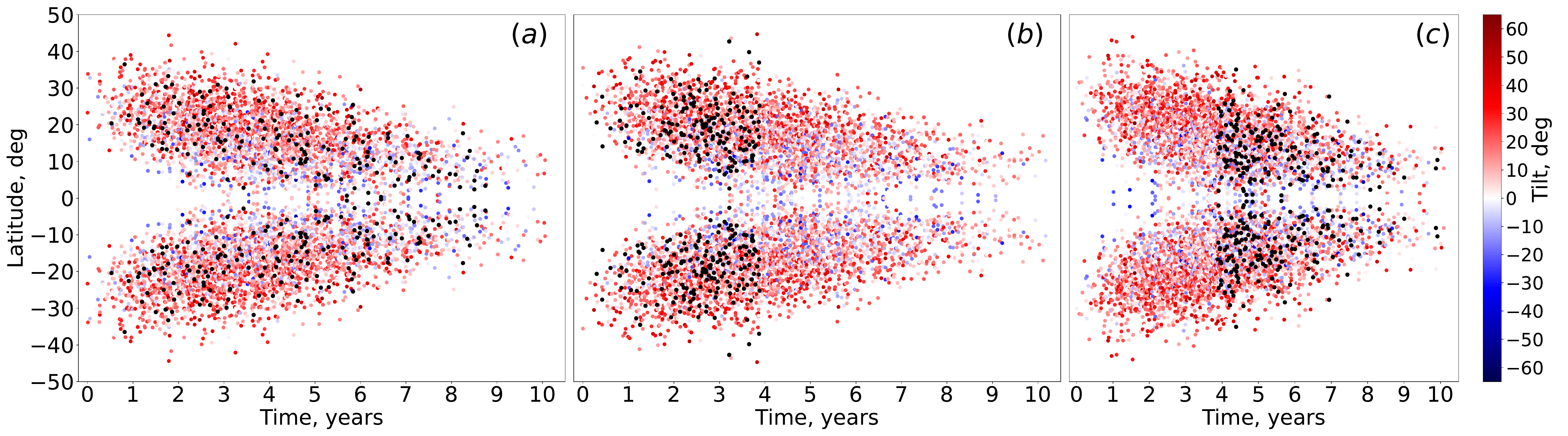} }
 \caption{The butterfly diagrams with the corresponding tilt information (colored dots). The red dots are the regular BMRs and the blue ones are the anti-Joy BMRs, while the black ones are the anti-Hale BMRs. The panels (a), (b), and (c) represents the Cases (iii), (iv), and (v) respectively.} 
 \label{bflys}
\end{figure*}

In this study, we present five different cases with different properties of the tilt of the BMRs. Here a list of these cases are given below:

\begin{itemize}
 \item[(i)]Cycles with all the BMR tilts obtained from the Joy's law.
 \item[(ii)] Cycles with variation in the BMR tilt properties (i.e. having anti-Joy BMRs), without any anti-Hale BMRs present in the cycles.
 \item[(iii)] Cycles with variation in the BMR tilt properties  and anti-Hale BMRs present throughout all the phases of the cycles.
 \item[(iv)] Cycles with variation in the BMR tilt properties and anti-Hale BMRs being present only in the rising  phases of the cycles.
 \item[(v)] Cycles with variation in the BMR tilt properties and anti-Hale BMRs being present only in the declining  phases of the cycles.
\end{itemize}

The last two cases are inspired by the fact that the polar field produced from the decay of BMRs during a significant part of the rising phase of a cycle is used to reverse the polar field, while during the solar maximum and the decline phase, the polar field is developed. Hence, the disturbance
in the rising phase of the cycle is expected to change the timing of the polar field reversal more significantly than the disurbances introduced during the decling phase \citep{Kitchatinov18, KMB18}.

In the \Fig{bflys} we present the typical butterfly diagrams of the input BMRs along with the information of their tilt as shown in the color scheme. The black dots shown here are the anti-Hale type BMRs where as the blue colored BMRs are of the anti-Joy type.  Here, the panel (a), (b), and (c) represents the Cases (iii), (iv), and (v) respectively. 

Here we mention that for the analysis of the reversal time of polar field, 30 cycles has been simulated for each of these cases. The amplitudes of the simulated solar cycles vary within a range of 30 to 90 in terms of monthly BMR number. In the cycles for the Cases (ii), (iii), (iv) and (v), the anti-Joy BMRs are present randomly throughout the cycles, whereas, the presence of the anti-Hale BMRs, in the last three cases, are varied in the different phases of the cycles as mentioned earlier.

Finally, in the \Table{rev}, we summarize the statistics of the timings of the polar field reversals for each of the above mentioned cases.
The values of the average reversal times of the polar field for different cases matches well with the results obtained from the observations as noted in the last column of \Table{res}.
From the \Table{rev}, it can be clearly seen that the presence of the `anomalous' BMRs in different phases of the cycles significantly impacts the reversal time. 
As in the Case (i) the tilts of the BMRs strictly follow the Joy's law, there are no `anomalous' regions present in this case. Hence the reversal time is the shortest for these cycles.
Comparing the Case (i) with the Case (ii), we can infer that the anti-Joy regions present in the cycles of the Case (ii) imposes only a slight delay in the reversal time. 
However, when the cycles of Case (iii) consists of some anti-Hale regions along with the anti-Joy regions, it produces a significant amount of delay in the reversal time. 
On the other hand, when the cycles of Case (iv) consists of  large amounts of anti-Hale regions concentrated in their initial phases, the delay is further enhanced to a large extent.
The rising phases of the cycles in Case (v) are very similar to those of the cycles in Case (ii), i.e. the cycles from both the cases have only anti-Joy BMRs in their initial phases. Hence, their reversals time also have similar values. However, it is worth mentioning that, on an average, the cycles of Case (v) take slightly more time for their polar field reversal than the cycles of Case (ii). This is due to the presence of significant amount of anti-Hale BMRs during the maxima phases of the cycles, which have caused significant delay in polar field reversal for few cycles of Case (v), making the average value of reversal time for this case to be slightly higher.

In this study, we have explored the possible cases for the presence of the anti-Hale regions keeping their amount to be within a range of 3-7\% for different cycles. As a result, when we spread them out throughout the cycle (Case (iii)), their temporal density is less compared to the case when they are present in a certain phase of the cycle (e.g., Case (iv)). In \Fig{bflys} for representation purpose, the amount of the anti-Hale BMRs have been kept the same (5\%) for all three cases.
We emphasize that, the significant delay in the polar field reversal can be caused due to an enhanced temporal density of anti-Hale BMRs in the beginning phases of the cycles as seen in Case (iv). However, when the temporal density of the anti-Hale BMRs for Case (iv) is kept similar to the Case (iii) by lowering their percentage amount, we would get a value of the reversal time similar to that of Case  (iii).

The results discussed above regarding the impact of anomalous regions on the polar field reversal time are in qualitative agreement with \citet{NLLPC17} and \citet{Pal23}. However,
our results indicate that, the anti-Hale regions are much greater source of disturbance in the evolution of the polar field than the anti-Joy regions.
The simulation results of Table~\ref{rev} further hints towards the possibility that the significant scatter in the observed values of $T_r$ for different cycles as presented in \Table{res} may have been caused due to the presence of anti-Hale and anti-Joy regions in varied amount throughout the different phases of the solar Cycles 21\,--\,24.

\begin{table}
 \caption{The mean of the reversal time ($<T_r>$) and their corresponding standard deviation ($\sigma_{T_r}$) from the simulations of 30 cycles belonging to each of the cases mentioned above. The unit of the quantities are presented in terms of years as well as in CRs.}
 \label{rev}
\centering
\begin{tabular}{lllllcl}
\cline{1-5}
Case &&  $<T_r> $  && $\sigma_{T_r}$  \\
~~~~~~~ && Years (CRs)   && Years (CRs)\\
\cline{1-5}
(i) && 3.59 (48.07) && 1.08 (14.46) \\
\cline{1-5}
(ii) && 3.67 (49.14) && 0.89 (11.92) \\
\cline{1-5}
(iii) && 3.85 (51.56) && 1.11 (14.86) \\
\cline{1-5}
(iv) && 4.19 (56.11) && 0.96 (12.85)\\
\cline{1-5}
(v) && 3.69 (49.41) && 1.11 (14.86) \\
\cline{1-5}
\end{tabular}
\end{table}

\section{Conclusion}
\label{Sect-Conclusion}  
To study the polar field reversals in Cycles 21\,--\,24, we considered temporal series of full-disk LoS-magnetograms from two ground-based and two space-based instruments. We analyzed them both in the original form and in the form after correction of every magnetogram for its zero-level offset in the polar caps. A value of the offset was determined collectively at the latitudes above $\pm55^\circ$, at Stonyhurst longitudes in the range of $\pm45^\circ$ from the central meridian and within heliocentrical distance of $87^\circ$. In both forms, magnetogram's LoS field values were converted to the radial component.

In a set of magnetograms for every CR, the magnetic field in the longitudinal range $\pm45^\circ$ from the central meridian was averaged over latitudinal zones to construct the time-latitude diagrams.
The diagrams led us to choose the working latitudinal ranges $\pm$(65\,--\,80)$^\circ$ to study the reversals of polar magnetic fields. The high-latitudinal zonally averaged values of the magnetic fields show significant seasonal variations.

We reduced the noise level of each diagram by removing from consideration the points where the magnetic field modulus was above a case-specific threshold. Then the variations of magnetic field averaged over the selected latitudes were considered. They were smoothed over 13~CRs to keep away the seasonal geometric effects.

The north polar zone was leading in the polarity reversals of the four considered cycles in the case of original data. Data corrected for zero-offset demonstrated synchronous north and south polarity reversals in Cycles 21\,--\,23 according to the MWO/STT, as well as the leadership of the north polar zone by 5\,--\,6~CRs in Cycle 24 according to the SDO/HMI and SOLIS/VSM. Avoiding extrapolations and mathematical excesses, we did not find a multiple polarity reversal in the considered Cycles.

Accurate measurements of the magnetic field at the poles are difficult due to the annual variations in the inclination of the Sun's rotation axis to the observer, due to known problems in magnetic zero-level offset, and due to instrumental noises, which are especially strong in the polar zones. Therefore, now it is impossible to determine time of a polar field reversal with high accuracy. Considering what was said in Section~\ref{Sect-Results}, we can draw a conclusion about the time limits for the singlet reversals considered at the north/south polar zones: CRs~1692\,--\,1695/1695\,--\,1700 ($T_r$ = 53\,--\,56/56\,--\,61) in Cycle~21, CRs~1825\,--\,1828/1828\,--\,1842 ($T_r$ = 46\,--\,49/49\,--\,63) in Cycle~22, CRs~1959\,--\,1965/1966\,--\,1968 ($T_r$ = 46\,--\,52/53\,--\,55) in Cycle~23, and CRs~2132\,--\,2137/2148\,--\,2153 ($T_r$ = 54\,--\,59/70\,--\,75) in Cycle~24. 

Our studies of the polar field reversal dynamics, under the assumption about the process uniformity, showed that a reversal at one of the poles lasts about 2 years, on average.
 
We performed SFT simulations, to probe the origin of the significant cycle-to-cycle variations in the polar field reversal time. We found that the variation in the tilt properties of BMRs and the presence of `anomalous' BMRs in different phases of the cycles impacts the polar field reversal time significantly. The presence of `anomalous' regions in the early phases of the cycles imparts a considerable delay in the reversal time. We also find the impact of anti-Hale regions on the reversal time to be more pronounced than the anti-Joy type BMRs.

\section*{Acknowledgements}
\addcontentsline{toc}{section}{Acknowledgements}
The work was financially supported by the Ministry of Science and Higher Education of the Russian Federation (G.E.M, Kh.A.I.). A.B. acknowledges the financial support provided by the University Grants Commission, Govt. of India.
BBK acknowledges financial support provided by Ramanujan Fellowship (project no SB/S2/RJN-017/2018).
We are very grateful to the anonymous reviewer for valuable comments that helped us improve the presentation of our results.

\section*{Data Availability}
This study includes data from the synoptic program at the 150-Foot Solar Tower of the Mt. Wilson Observatory. The Mt. Wilson 150-Foot Solar Tower is operated by UCLA, with funding from NASA, ONR and NSF, under agreement with the Mt. Wilson Institute.
Data were acquired by SOLIS instruments operated by NISP/NSO/AURA/NSF. 
SOHO is a project of international cooperation between ESA and NASA. 
Data are available by courtesy of NASA/SDO and the AIA, EVE, and HMI science teams.
Sunspot data from the World Data Center SILSO, Royal Observatory of Belgium, Brussels.

\bibliographystyle{mnras}
\bibliography{Paper}

\label{lastpage}
\end{document}